# Preference-Aligned Options from Generative AI Compensates for Age-Related Cognitive Decline in Decision Making

Sayaka Ishibashi, Kou Tamura, Ayana Goma, Kenta Yamamoto, Kouhei Masumoto

Graduate School of Human Development and Environment, Kobe University, Japan

Correspondence concerning this article should be addressed to Kouhei Masumoto, Graduate School of Human Development and Environment, Kobe University, 3-11 Tsurukabuto, Nada-ku, Kobe, Hyogo 657-8501, Japan. TEL: +81-78-803-7892. E-mail: masumoto@people.kobe-u.ac.jp

## Abstract

Older adults often experience increased difficulty in decision making due to age-related declines particularly in contexts that require information search or the generation of options from memory. This study examined whether using generative AI for information search enhances choice satisfaction and reduces choice difficulty among older adults. A total of 130 participants (younger, *n* = 56; older, *n* = 74) completed a music-selection task under AI-use and AI-nonuse conditions across two contexts: previously experienced (road trip) and not previously experienced (space travel). In the AI-nonuse condition, participants generated candidate options from memory; in the AI-use condition, GPT-4o presented options tailored to individual preferences. Cognitive functions, including working memory, processing speed, verbal comprehension, and perceptual reasoning, were assessed. Results showed that AI use significantly reduced perceived choice difficulty across age groups, with larger benefits in unfamiliar contexts. Regarding cognitive function, among older adults, lower cognitive function was associated with fewer recalled options, higher choice difficulty, and lower satisfaction in the AI-nonuse condition; these associations were substantially attenuated when AI was used. These results demonstrate that generative AI can mitigate age-related cognitive constraints by reducing the cognitive load associated with information search during decision making. While the use of AI reduced perceived difficulty, choice satisfaction remained unchanged, suggesting that autonomy in decision making was preserved. These findings indicate that generative AI can support everyday decision making by compensating for the constraints in information search that older adults face due to cognitive decline.

# 1. Introduction

Decision making in later life plays a central role in determining well-being and independence, influencing judgments about continued employment, medical treatment, financial risk management, and long-term life planning. At the same time, aging is accompanied by declines in several cognitive domains, including memory retrieval, processing speed, working memory, and executive function, which in turn make it more difficult for older adults to manage decisions involving complex alternatives or decisions under uncertainty, to respond rapidly, and to engage in novel decision contexts or information search (Mather, 2006; Peters et al., 2007). Consequently, older adults may experience diminished decision confidence and avoid demanding decisions to reduce immediate stress, a strategy that, while adaptive in the short term, may lead to long-term disadvantages (Mather, 2006). Against this backdrop, supporting decision making in later life has become an increasingly important research and societal priority. Because decision goals, values, and preferences differ widely among individuals (Morelli et al., 2022), personalized decision support remains difficult to achieve. Generative AI based on large language models (LLMs) now offers the capacity to generate contextually relevant alternatives in real time for complex, highly individualized queries by drawing on vast information sources. As a result, the potential of LLM-enabled systems to assist older adults in making informed and preference-consistent decisions has attracted growing research and practical interest (Abadir et al., 2024; Wang & Lighthall, 2025). Shandilya and Fan (2022) conducted qualitative interviews with older adults who use AI-enabled technologies, including recommendation systems on YouTube and Netflix, online shopping platforms, and voice assistants such as Alexa. The study revealed that many older adults view AI as a supportive tool that facilitates everyday activities. However, systematic research examining how generative AI can compensate for age-related cognitive decline, and its impact on decision satisfaction and perceived choice difficulty, remains limited.

Episodic memory and working memory decline with age (Park et al., 2002; Salthouse, 2019), and these declines have been reported to be associated with greater difficulty in retrieving solutions to personal problems (Peters et al., 2019). Age-related decline in working memory further limits the ability to maintain and compare multiple alternatives, thereby constraining information search during decision making (Mather, 2006). Consequently, in decision-making tasks that require reference to one's own past experiences, age-related declines in episodic and working memory may impose fundamental constraints on experience-based decision processes among older adults. This study aimed to experimentally investigate whether, during the information-search phase of decision making, generative AI can enhance choice satisfaction and reduce perceived choice difficulty among older adults by presenting candidate options tailored to individual preferences. Mikels et al. (2010) compared an emotion-focused condition (in which participants chose by attending to their affective reactions to options), an information-focused condition (in which participants chose based on option attributes), and a control condition, demonstrating that choice quality was highest among older adults in the emotion-focused condition, whereas younger adults performed best in the information-focused condition. In the present study, we prompted the generative AI to generate emotion-focused, preference-aligned option sets and examined their effects on decision evaluations. Accordingly, we hypothesized that, by presenting preference- aligned options and reducing the cognitive load involved in information search, generative AI would reduce perceived choice difficulty and increase choice satisfaction among older adults to a greater extent than among younger adults.

The second aim of this study was to examine whether the usefulness of AI differs between unfamiliar choices and choices that have been previously experienced. Prior research indicates age-related differences in information processing during decision making: younger adults tend to rely on more analytic and deliberative processes, whereas older adults are more likely to draw on experience-based, intuitive, and affective processes (e.g., Li et al., 2013; Peters et al., 2007). Accordingly, among younger adults, the options presented by generative AI make it possible to allocate cognitive resources that would be required for information search to more deliberative processes for judgment, thereby increasing choice satisfaction and reducing perceived choice difficulty. By contrast, for older adults, the effects of generative AI use are expected to be limited in decision contexts that are already familiar, because they can draw on their own prior experiences. However, in unfamiliar decision contexts, presenting preference-aligned options through generative AI is expected to alleviate cognitive load, leading to lower perceived choice difficulty and higher choice satisfaction.

We also examined associations between cognitive function and both choice satisfaction and choice difficulty. Age-related declines in cognitive function are expected to affect information search and the comparison of choice alternatives (Mather, 2006; Peters et al., 2007), thereby increasing perceived choice difficulty and reducing choice satisfaction. However, when generative AI is used, preference-aligned option recommendations are expected to compensate for these declines; accordingly, the associations between lower cognitive function and the evaluative outcomes of decision making should be attenuated or eliminated under AI-use conditions.

## 2. Methods

### *2.1. Experimental Design*

The study employed a three-factor mixed design with Age group (younger, older), AI condition (AI-use, AI-nonuse), and Experience condition (road trip, space travel). Age group was treated as a between-participants factor, whereas AI condition and Experience condition were within-participants factors.

### *2.2. Participants*

The experimenters invited 200 individuals who had participated in previous studies, and 130 of them consented to take part in the present experiment. The younger group consisted of 56 participants (21 men and 35 women; M = 21.34, SD = 1.78), and the older group consisted of 74 participants (38 men and 36 women; M = 74.23, SD = 5.06). A priori power analysis for a mixed design with two within-participants factors (AI condition, Experience condition) and one between-participants factor (age group) indicated that a sample size of 41 participants per age group (82 in total) would be required to detect a moderate effect size ($f = .25$) with 80% power (Faul et al., 2007). Therefore, the sample size in the present study can be considered adequate. None of the participants had a history of neurological or psychiatric disorders. Cognitive function was assessed using the Wechsler Adult Intelligence Scale–Fourth Edition (WAIS-IV; Wechsler, 2008). Based on ten core subtests, the Full Scale IQ (FSIQ), Verbal Comprehension Index (VCI), Perceptual Reasoning Index (PRI), Working Memory Index (WMI), and Processing Speed Index (PSI) were calculated. The FSIQ was within the normal range ($M = 115.84$, $SD = 14.95$) and no participants showed indications suggestive of dementia. Results of between-group comparisons on the 10 subtests of the WAIS-IV indicated that the younger group scored significantly higher than the older group on all subtests except Vocabulary ($t$ (128) = −1.08, p

= .28; *ps* < .01 for all other subtests). Vocabulary is one of the subtests that assesses crystallized knowledge, which tends to be preserved in later life. Therefore, the cognitive profile of the older adults in this study can be considered typical of age-related patterns of cognitive performance.

### *2.3. Decision-Making Task*

The decision-making task was developed with reference to the AI-based music recommendation experiment (Jin et al., 2019). In the present study, two experience conditions were established to examine the effect of prior decision-making experience on choice satisfaction and difficulty. In the *road trip* condition, representing a familiar decision-making context, participants were asked to select one song they would like to listen to during a "road trip." In the *space travel* condition, representing an unfamiliar context, participants were asked to select one song they would like to listen to during a "space travel inside a spacecraft." The order of the experience conditions was counterbalanced across participants.

Two conditions of music selection were established: the AI-nonuse condition and the AI-use condition. Participants first completed the AI-nonuse condition, followed by the AI-use condition. In the *AI-nonuse condition*, participants were instructed to recall as many songs as possible that fit the given experience condition and to write them down on paper. They were prohibited from using the internet and were required to generate candidate songs solely from memory. The time limit was set at five minutes. Subsequently, participants were asked to select the single the single most suitable song from their recalled list. In the *AI-use condition*, GPT-4o was employed for music selection. Prior to this condition, participants completed a brief survey regarding their musical preferences (Appendix A), including preferred genres, eras, moods, and tempos. Based on each participant's responses, the generative AI was instructed to present five candidate songs tailored to the participant’s musical preferences, together with links to the corresponding YouTube music videos, so that the participant could review each option. The AI was operated by the experimenter, and the proposed songs were displayed on a monitor in front of the participant. Participants were then asked to choose the song they found most suitable from among the AI-generated options. If none of the proposed songs matched their preferences, the AI was prompted to generate additional candidates. They were required to make a selection within five minutes.

After completing both the AI and AI-nonuse conditions, participants evaluated their choice satisfaction and choice difficulty. Choice satisfaction was rated on a 5-point Likert scale (1 = “Not at all satisfied” to 5 = “Very satisfied”), and choice difficulty was also rated on a 5-point Likert scale (1 = “Not at all difficult” to 5 = “Very difficult”). In addition, participants were asked about their prior experience with generative AI use and their background in music.

### *2.4 Procedure*

The experiment was conducted individually with each participant. The experiment was carried out over two separate days: on the first day, participants completed the WAIS-IV, and on the second day, they performed the decision-making task.

### *2.5 Ethical considerations*

This study was conducted with the approval of the research ethics committee of the author’s affiliated institution (no.750). Written informed consent was obtained from all individual participants included in the study.

## 3. Results

### *3.1. Number of Recalled Songs*

Figure 1 presents the number of songs recalled by participants in the AI-nonuse condition, classified by age group and experience condition. A two-way ANOVA on the number of recalled songs revealed significant main effects of experience condition( $F$ (1, 128) = 44.56, $p < .001$, $\eta^2_p = .26$), and age group ($F$ (1, 128) = 24.62, $p < .001$, $\eta^2_p = .16$), as well as a significant interaction between these factors ($F$ (1, 128) = 6.98, $p = .01$, $\eta^2_p = .05$). Given the significant interaction, tests of simple main effects were conducted. The results indicated that, regardless of experience condition, older adults recalled significantly fewer songs than younger adults (road trip condition: $F$ (1, 128) = 25.49, $p < .001$, $\eta^2_p = .17$; space travel condition: $F$ (1, 128) = 15.00, $p < .001$, $\eta^2_p = .11$). The age difference was larger in the road trip condition than in the space travel condition. Additionally, in both age groups, participants recalled significantly more songs in the road trip condition than in the space travel condition (younger group: $F$ (1, 128) = 38.12, $p < .001$, $\eta^2_p = .23$; older group: $F$ (1, 128) = 9.45, $p < .001$, $\eta^2_p = .07$), with the difference between the two experience conditions being greater in the younger group than in the older group.

Overall, these results indicate that the older group recalled fewer songs than the younger group, and that younger group exhibited greater experience-based differences in the number of recalled songs.

### *3.2. Choice satisfaction*

Figure 2 illustrates choice satisfaction as a function of age group, AI condition (use vs. non-use), and experience condition (road trip vs. space travel). A three-way ANOVA revealed a significant main effect of experience condition ($F$ (1, 128) = 14.05, $p < .001$, $\eta^2_p = .10$).

Participants reported higher satisfaction with their music selection in the *road trip* condition compared to the *space travel* condition. There were no significant main effects of the AI condition or age group (AI condition: $F$ (1, 128) = 0.93, $p$ = .34, $\eta^2_p$ = .01; age group: $F$ (1, 128) = 0.92, $p$ = .34, $\eta^2_p$ = .01) and no significant interactions (AI condition × age group, $F$ (1, 128) = 0.19, $p$ = .66, $\eta^2_p$ = .00; experience condition × age group, $F$ (1, 128) = 1.40, $p$ = .24, $\eta^2_p$ = .01; AI condition × experience condition, $F$(1, 128) = 0.81, $p$ = .37, $\eta^2_p$ = .01; three-way interaction, $F$ (1, 128) = 0.17, $p$ = .69, $\eta^2_p$ = .00).

### *3.3 Choice difficulty*

Figure 3 shows choice difficulty. A three-way ANOVA was conducted with choice difficulty as the dependent variable. A significant main effect of the AI condition was observed, ($F$ (1, 128) = 48.59, $p < .001$, $\eta^2_p$ = .28). Participants in the AI-use condition reported significantly lower choice difficulty compared to those in the AI-nonuse condition. A significant main effect of the experience condition was also found ($F$ (1, 128) = 43.60, $p < .001$, $\eta^2_p$ = .25), indicating that selecting music for the unfamiliar experience (space travel condition) was associated with higher choice difficulty than for the familiar experience (road trip condition). The main effect of age group was not significant ($F$ (1, 128) = 1.17, $p$ = .28, $\eta^2_p$ = .01). Significant interactions were found for AI condition × age group ($F$ (1, 128) = 4.44, $p$ = .04, $\eta^2_p$ = .03) and AI condition × experience condition ($F$ (1, 128) = 6.74, $p$ = .01, $\eta^2_p$ = .05). In contrast, the interaction between experience condition × age group was not significant ($F$ (1, 128) = 2.40, $p$ = .12, $\eta^2_p$ = .02), nor was the three-way interaction ($F$ (1, 128) = 0.01, $p$ = .92, $\eta^2_p$ = .00). Post-hoc analyses for the AI condition × age group interaction revealed that, in both age groups, choice difficulty was significantly lower in the AI-use condition compared to the AI-nonuse condition (both *ps* $< .001$). However, in the AI-use condition, older adults reported significantly higher choice difficulty than younger adults ($p = .04$). Regarding the AI condition × experience condition interaction, regardless of experience condition, participants in the AI-nonuse condition reported higher difficulty than those in the AI-use condition (*ps* $< .001$ for both experience condition). The difference between AI-use and AI-nonuse condition was larger for the space travel condition than for the road trip condition. Moreover, choice difficulty was higher for space travel condition than for road trip condition in both AI conditions (*ps* $< .001$), although this difference was more pronounced in the AI-nonuse condition.

### *3.4. Correlations Among Variables*

Table 1 and 2 present the results of Pearson's correlation analyses among the number of recalled songs, choice satisfaction, choice difficulty, and cognitive function tests for the road trip and space travel conditions, respectively, separately for the older and younger groups.

#### *3.4.1. Associations Among the Number of Recalled Songs, Choice Satisfaction, and Choice Difficulty*

In the AI-nonuse condition, older adults showed that a greater number of recalled songs was associated with higher choice satisfaction and lower choice difficulty in both the driving (choice satisfaction, $r = .26$, $p = .02$; choice difficulty, $r = -.34$, $p < .01$) and space-travel conditions (choice satisfaction, $r = .23$, $p = .05$; choice difficulty, $r = -.24$, $p = .04$). In contrast, under the AI-use condition, no significant correlations were observed between the number of recalled songs and either choice satisfaction or choice difficulty.

Among younger adults in the AI-nonuse condition, no significant correlations were found between the number of recalled songs and either choice satisfaction or choice difficulty in the road trip condition. In the AI-use condition, a greater number of recalled songs was associated with lower levels of choice satisfaction in the space travel condition ($r = -.34$, $p = .01$). Regarding the relationship between choice satisfaction and choice difficulty, a significant negative correlation was observed in all conditions except for younger adults in the AI-nonuse/space travel condition.

### *3.4.2. Associations with Cognitive Function*

**Road Trip Condition**： Among older adults, a greater number of recalled songs was associated with higher cognitive function (FSIQ, $r = .24$, $p = .04$; VCI, $r = .25$, $p = .03$; PRI, $r = .24$, $p = .04$). No significant correlations were observed between choice satisfaction and cognitive function. Regarding choice difficulty, in the AI-nonuse condition, higher levels of perceived difficulty were associated with lower cognitive function (FSIQ, $r = -.29$, $p = .01$; VCI, $r = -.30$, $p = .01$; PRI, $r = -.25$, $p = .03$; WMI, $r = -.25$, $p = .03$). However, in the AI-use condition, no significant correlations were found between cognitive function and choice difficulty.

Among younger adults, no significant correlations were found between the number of recalled songs and cognitive function. Regarding choice satisfaction, no significant correlations with cognitive function were observed in AI-nonuse condition; however, in the AI-use condition, higher PSI was associated with higher choice satisfaction ($r = .33$, $p = .01$). For choice difficulty, higher VCI were associated with higher choice difficulty in both the AI-nonuse ($r = .37$, $p = .01$) and AI-use conditions ($r = .28$, $p = .04$).

**Space Travel Condition:** Among older adults, no significant correlations were observed between the number of recalled songs and cognitive function. Regarding choice satisfaction, in the AI-nonuse condition, higher cognitive function was associated with higher satisfaction (FSIQ, $r$ = .29, $p$ = .01; VCI, $r$ = .26, $p$ = .02; WMI, $r$ = .24, $p$ = .04; PSI, $r$ = .28, $p$ = .01); however, in the AI-use condition, no significant correlations with cognitive function were found. For choice difficulty, older adults with higher VCI reported lower choice difficulty in both the AI-nonuse (VCI, $r$ = -.23, $p$ = .05) and AI-use (VCI, $r$ = -.27, $p$ = .02) conditions.

Among younger adults, no significant correlations were found between the number of recalled songs and cognitive function. In terms of choice satisfaction, in the AI-nonuse condition, higher cognitive function (FSIQ, $r$ = -.28, $p$ = .04; PRI, $r$ = -.28, $p$ = .04) were associated with lower choice satisfaction, whereas in the AI-use condition, lower VCI was associated with higher satisfaction ($r$ = -.34, $p$ = .01). Conversely, no significant correlations were observed between cognitive function and choice difficulty in either the AI-use or AI-nonuse conditions.

## 4. Discussion

The purpose of this study was to examine whether choice options generated by AI based on individual preferences enhance choice satisfaction and reduce choice difficulty among older adults, and to clarify whether the influence of generative AI differs between previously experienced choices and new choices. The results revealed the following findings. 1) Regarding the number of recalled song options, older adults recalled fewer songs than younger adults in the AI-nonuse condition. In both age groups, the number of recalled songs was smaller for unfamiliar choice condition (space travel) than for familiar choice condition (road trip). 2) For choice satisfaction, no significant effect of AI use was observed. Only the main effect of experience was significant, indicating that participants reported greater satisfaction in the familiar choice condition than in the unfamiliar choice condition. 3) Regarding choice difficulty, both older and younger adults showed reduced perceived difficulty when using AI. However, older adults reported greater choice difficulty than younger adults even when using AI. 4) With respect to cognitive function, among older adults, in the AI-nonuse condition, lower cognitive function was associated with recalling fewer songs, lower choice satisfaction, and higher choice difficulty. In contrast, in the AI-use condition, the associations between cognitive function and both choice satisfaction and choice difficulty were substantially attenuated.

### *4.1. Number of Recalled Song Options, Choice Satisfaction, and Choice Difficulty in the AI-nonuse Condition*

In the AI-nonuse condition, older adults recalled fewer song options than younger adults in both the road trip and space travel conditions. Previous studies have similarly reported that information search decreases with age (Mather, 2006). Furthermore, a significant correlation was found between the number of recalled songs and cognitive function in this study, suggesting that the smaller number of recalled songs among older adults may be attributable to age-related declines in memory retrieval and recall ability (Park et al., 2002; Salthouse, 2019), which may make it more difficult for them to retrieve songs appropriate to the given context from their existing knowledge base. In addition, among older adults fewer recalled songs were associated with lower choice satisfaction and higher choice difficulty. There are important links between memory and decision-making, and individuals can improve future decisions by referencing past experiences (Biderman et al., 2020). These findings indicate that age-related declines in information search ability during decision-making may negatively affect subjective evaluations of one's own decisions.

Among younger adults, no significant correlations with choice satisfaction and difficulty emerged in either the space travel or road trip condition. Evaluation of decision outcomes such as choice satisfaction and regret are more strongly influenced by factors including decision justification (Reb & Connolly, 2010) and self-efficacy (Pignault et al., 2023). It is therefore possible that, for younger adults who do not exhibit age-related cognitive declines, choice satisfaction and difficulty were influenced less by the sheer number of recalled options and more by their ability to articulate convincing reasons for their choices and by their belief in their own decision-making competence.

### *4.2. Changes in Choice Satisfaction and Choice Difficulty with AI Use*

This study showed that the use of AI reduced perceived choice difficulty for both older and younger adults. It is conceivable that by presenting candidate songs based on participants' musical preferences, the generative AI reduced the cognitive load associated with searching for potential options, thereby contributing to the observed decrease in perceived choice difficulty. Regarding age-group differences, we predicted that older adults, who tend to prioritize the emotion evoked by the options (Mikels et al., 2010), would benefit more than younger adults from preference-aligned information provided by the AI. However, contrary to this prediction, the results showed that the reduction in choice difficulty was greater for younger adults than for older adults. A plausible explanation is that comparing and selecting the single most appropriate song from the five AI-generated options may have imposed additional cognitive demands on older adults, thereby diminishing the benefit. Prior research suggests that older adults tend to prefer situations in which fewer options are presented and that they perform optimally under such conditions (Reed et al., 2008). For younger adults, evaluating five alternatives may not have imposed sufficient cognitive demands to affect perceived choice difficulty. Another possibility is that effective decision support for older adults depends on transparent communication that fosters full comprehension of provided information (American Psychological Association, 2024). Thus, the black-box nature of generative AI may have impaired comprehension and acceptance, increased additional psychological strain, and consequently led to a smaller reduction in perceived choice difficulty among older adults relative to younger adults. Moreover, perceived usefulness of and trust in AI influence older adults' acceptance of, and intention to use, AI-based technologies (Shandilya & Fan, 2022). Therefore, age-related differences in these perceptions may partly explain why the facilitative effect of AI on choice difficulty was attenuated in older adults relative to younger adults.

With respect to prior decision experience, for both age groups, the AI-induced reduction in choice difficulty was larger in the novel context (space travel) than in the previous experienced context (road trip). When individuals perceive low subjective knowledge, they tend to anticipate greater decision difficulty, and in such situations, exposure to a larger set of options can enhance self-efficacy, defined as the belief that one can make a good choice (Hadar & Sood, 2014). The present findings suggest that generative AI is particularly beneficial in novel decision contexts, as it can present multiple preference-aligned options that compensate for a lack of knowledge and enhance decision-related self-efficacy.

With respect to choice satisfaction, neither younger nor older adults showed significant changes as a function of AI use. Although greater choice difficulty typically predicts lower satisfaction (Chernev et al., 2015), the two constructs are influenced by partially distinct determinants. Choice difficulty is primarily affected by task complexity and the availability of cognitive resources (Lv et al., 2023; Misuraca et al., 2024), whereas choice satisfaction is more closely tied to self-related factors, including perceived autonomy and involvement in the decision process (Birkeland et al., 2022; Botti & Iyengar, 2006; Botti et al., 2023). In this study, the AI did not present a fixed set of predetermined options; instead, it generated multiple preference-aligned options from which participants selected the single option they considered most appropriate. This interactive structure likely maintained participants' sense of agency and perceived autonomy in the decision process, which may, in turn, explain why AI use did not alter choice satisfaction.

Additionally, among older adults, lower cognitive function under the AI-nonuse condition was associated with lower choice satisfaction and higher perceived choice difficulty, whereas these associations largely disappeared under AI-use conditions, except for a negative correlation between VCI and choice difficulty. This pattern suggests that AI support may have compensated for age-related declines in cognitive processing demands. Because the VCI reflects verbal knowledge, which may facilitate the evaluation and comparison of AI-generated alternatives, individuals with higher VCI may continue to experience lower perceived difficulty even when supported by AI.

### ***4.3.Limitations***

Empirical research examining how the use of generative AI in decision-making contexts affects choice satisfaction and choice difficulty among older adults remains limited, and, to the best of our knowledge, no prior study has explored these outcomes in relation to cognitive functioning. In addition to addressing these gaps, the present study is the first to examine whether the effects of generative AI differ depending on prior decision experience.

Despite these contributions, this study has several limitations. Although the music-selection task allowed a controlled comparison of familiar and novel decision contexts across age groups, the characteristics of this task made clear instance of "failed choices" unlikely, and choice satisfaction remained relatively high even without AI support. Consequently, the effect of generative AI on choice satisfaction may have been underestimated. Future research should examine the utility of AI-generated options in decision contexts where the potential for failure or outcome uncertainty is more pronounced. Moreover, this study examined whether AI can compensate for age-related declines in cognitive functioning that limit information search. However, previous research suggests that in highly self-relevant decisions, such as treatment choices, or in situations involving personal accountability and goal achievement, the impact of age-related cognitive decline on information search during decision making may differ (Liu et al., 2021; Strough et al., 2015). Future research should examine how AI generated options with adapted to individuals' goals and transparent explanations influence choice satisfaction and difficulty.

## 5. Conclusion

This study examined how the use of generative AI during decision making among older adults affects choice satisfaction and choice difficulty. Without AI support, age-related declines in cognitive function constrained information search and increased choice difficulty. By contrast, when generative AI presented options tailored to individual preferences, these constraints were partially compensated for, reducing choice difficulty. In particular, because older adults tend to experience greater difficulty as decision contexts become more novel, the observed reduction in choice difficulty resulting from AI-facilitated information search in unfamiliar contexts demonstrates the potential utility of generative AI in supporting decision making among older adults. However, no change in choice satisfaction was observed with AI use. This finding suggests that preference-aligned information search supported by generative AI does not compromise older adults' desirable choices. Previous research indicates that choice satisfaction is shaped by perceived control over the decision process and the extent of self-involvement (Botti & Iyengar, 2006). Accordingly, enhancing satisfaction through AI may require design approaches that actively foster user involvement and autonomy. Moreover, the reduction in choice difficulty with generative AI use was greater among younger adults than older adults. This pattern may reflect older adults' lower familiarity with AI technologies and heightened concerns regarding data privacy and the handling of personal information (Mathur et al., 2025). These findings suggest the need for educational interventions designed to enhance AI literacy and promote confident and informed use of AI systems among older adults.

**Declaration of Interests**

The authors declare that they have no known competing financial interests or personal relationships that could have appeared to influence the work reported in this paper.

**Declaration of Generative AI use**

We used ChatGPT-5 for proofreading, grammatical improvement, and translation. After using these tools, the authors reviewed and edited the content as needed and take full responsibility for the final version of the manuscript.

**Funding**

This work was supported by a Grant-in-Aid for JSPS (Japan Society for the Promotion of Science) KAKENHI (grant numbers 23K22352, 22H00078, 22H00088).

**Table 1**

Pearson Correlation Analysis in the Unfamiliar Decision Context (Road Trip).

| | | 1 | 2 | 3 | 4 | 5 | 6 | 7 | 8 | 9 | 10 |
|---|---|---|---|---|---|---|---|---|---|---|---|
| 1 | Number of recalled songs in the AI- nonuse condition | – | .04 | -.23 | .16 | .04 | .15 | -.09 | .14 | .12 | .15 |
| 2 | Choice satisfaction in the AI-nonuse condition | .26* | – | -.33* | .27* | -.12 | -.07 | -.16 | -.02 | .07 | -.06 |
| 3 | Choice difficulty in the AI-nonuse condition | -.34** | -.27* | – | -.01 | .21 | .12 | .37** | -.21 | .05 | .09 |
| 4 | Choice satisfaction in the AI-use condition | -.10 | .02 | .01 | – | -.44** | .12 | -.02 | -.03 | .02 | .33* |
| 5 | Choice difficulty in the AI-use condition | -.03 | -.03 | .37** | -.37** | – | .06 | .28* | -.10 | .02 | -.09 |
| 6 | FSIQ | .24* | .11 | -.29* | -.03 | -.06 | – | .63** | .57** | .66** | .43** |
| 7 | VCI | .25* | .06 | -.30* | -.09 | -.08 | .78** | – | .12 | .20 | .18 |
| 8 | PRI | .24* | .08 | -.25* | -.11 | .02 | .86** | .56** | – | .29* | -.18 |
| 9 | WMI | .14 | .08 | -.25* | .06 | -.10 | .81** | .49** | .62** | – | .14 |
| 10 | PSI | .16 | .14 | -.11 | .09 | -.02 | .73** | .37** | .53** | .53** | – |

*Note.* Upper-right cells show correlations for younger adults; lower-left cells show correlations for older adults; FSIQ = Full Scale Intelligence Quotient; VCI = Verbal Comprehension Index; PRI = Perceptual Reasoning Index; WMI = Working Memory Index; PSI = Processing Speed Index. *$p < .05$, **$p < .01$.

**Table 2**

Pearson Correlation Analysis in the Unfamiliar Decision Context (Space Travel).

| | | 1 | 2 | 3 | 4 | 5 | 6 | 7 | 8 | 9 | 10 |
|---|---|---|---|---|---|---|---|---|---|---|---|
| 1 | Number of recalled songs in the AI- nonuse condition | – | .15 | -.12 | -.34$^{**}$ | .21 | .25 | .11 | .09 | .12 | .26 |
| 2 | Choice satisfaction in the AI-nonuse condition | .23$^{*}$ | – | -.18 | -.01 | -.06 | -.28$^{*}$ | -.19 | -.28$^{*}$ | -.19 | .05 |
| 3 | Choice difficulty in the AI-nonuse condition | -.24$^{*}$ | -.34$^{**}$ | – | .02 | -.06 | .02 | .14 | .05 | -.20 | .00 |
| 4 | Choice satisfaction in the AI-use condition | .06 | .31$^{**}$ | .04 | – | -.42$^{**}$ | -.09 | -.34$^{*}$ | .13 | .03 | -.01 |
| 5 | Choice difficulty in the AI-use condition | -.11 | -.39$^{**}$ | .36$^{**}$ | -.24$^{*}$ | – | .10 | .15 | .01 | .04 | .01 |
| 6 | FSIQ | .14 | .29$^{*}$ | -.15 | .09 | -.14 | – | .66$^{**}$ | .57$^{**}$ | .66$^{**}$ | .43$^{**}$ |
| 7 | VCI | .15 | .26$^{*}$ | -.23$^{*}$ | .05 | -.27$^{*}$ | .78$^{**}$ | – | .12 | .20 | .18 |
| 8 | PRI | .08 | .16 | .01 | .03 | -.04 | .86$^{**}$ | .56$^{**}$ | – | .29$^{*}$ | -.18 |
| 9 | WMI | .04 | .24$^{*}$ | -.15 | .08 | -.08 | .81$^{**}$ | .49$^{**}$ | .62$^{**}$ | – | .14 |
| 10 | PSI | .20 | .28$^{*}$ | -.10 | .17 | -.01 | .73$^{**}$ | .37$^{**}$ | .53$^{**}$ | .53$^{**}$ | – |

*Note.* Upper-right cells show correlations for younger adults; lower-left cells show correlations for older adults; FSIQ = Full Scale Intelligence Quotient; VCI = Verbal Comprehension Index; PRI = Perceptual Reasoning Index; WMI = Working Memory Index; PSI = Processing Speed Index. $^{*}p < .05$, $^{**}p < .01$.

**Figure 1**. Mean  number of recalled songs in the AI nonuse condition

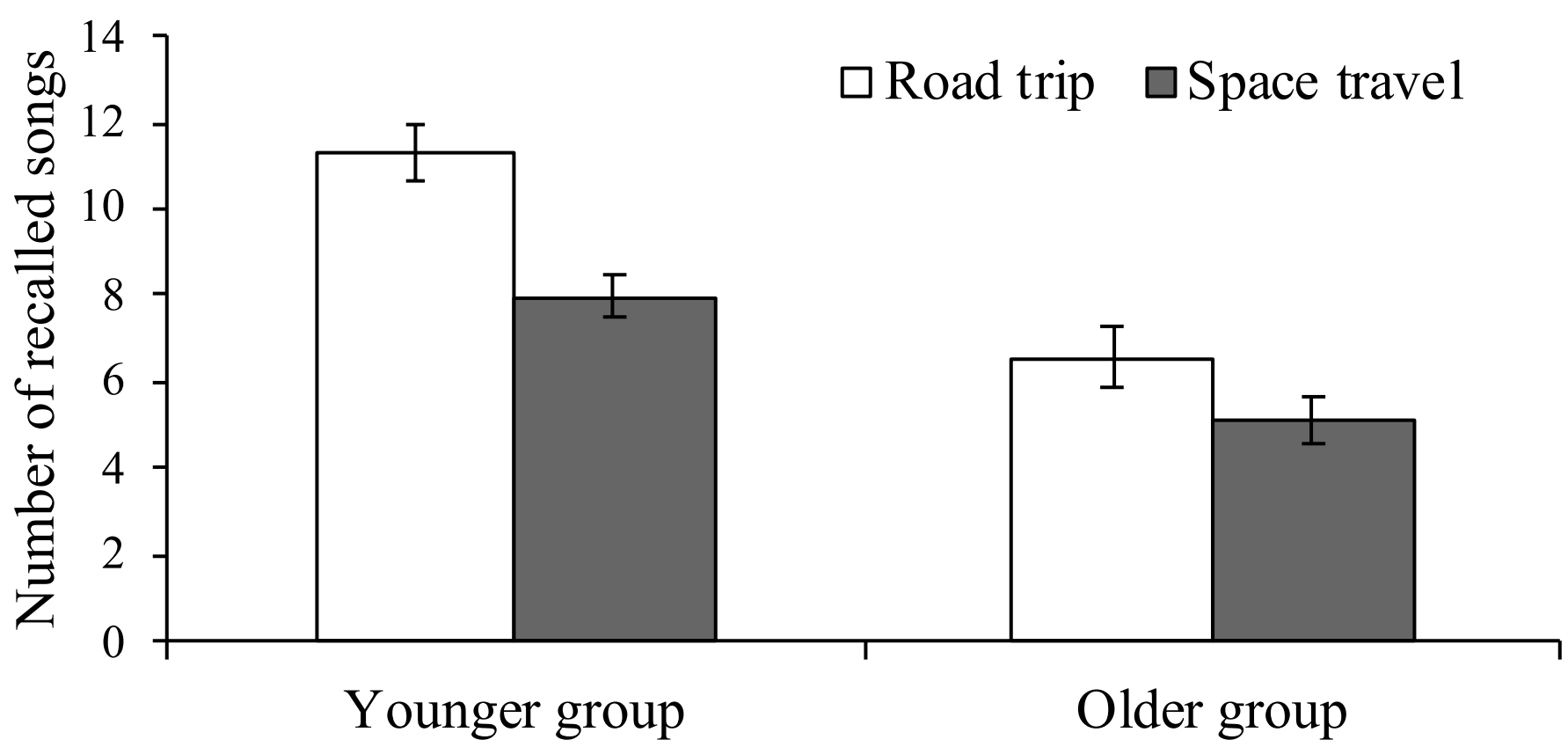

*Note*. Error bars represent standard errors.

**Figure 2**. Choice Satisfaction Across Age Groups, AI Conditions, and Experience Conditions

Score

5
4.5
4
3.5
3
2.5
2
1.5
1
0.5
0

Road trip | Space travel | Road trip | Space travel | Road trip | Space travel | Road trip | Space travel

AI-nonuse | AI-use | AI-nonuse | AI-use

Younger group | Older group

Note. Error bars represent standard errors.

**Figure 3.** Choice Difficulty Across Age Groups, AI Conditions, and Experience Conditions

Score

5
4.5
4
3.5
3
2.5
2
1.5
1
0.5
0

Road trip | Space travel | Road trip | Space travel | Road trip | Space travel | Road trip | Space travel

AI-nonuse | AI-use | AI-nonuse | AI-use

Younger group | Older group

**Appendix A:** Music Preference Checklist

## Music I'd like to listen to on a road trip

**1) Please choose your favorite genres form the list below. (You may select multiple.)**

Popular Songs / Hip Hop / Rock / Anime Songs / Vocaloid / Western Music / Classical / Jazz / Alternative / Blues / Country / Dance / R&B / Enka / Electronic / Soundtrack / Metal / K-POP / Latin

Other ( )

**2) Please select your preferred era (release period).**

～ 1970s / 1980s / 1990s / 2000s / 2010s / 2020s ～

**3) Please choose your preferred mood or tone of music. (You may select multiple.)**

Bright / Dark / Bittersweet / Fun / Humorous / Lively / Calm / Cool / Intense / Sad / Grand / Noisy / Warm / Sweet / Beautiful / Complex / Clear / Clean / Rebellious / Mature / Light & Rhythmical

Other ( )

**4) Please choose your preferred tempo (beats per minute).**

Slow Tempo ♩＝60–90

Medium Tempo (moderate speed) ♩＝100–110

Up-Tempo ♩＝110–140

Very Fast Tempo ♩＝150–200

Extremely Fast / Aggressive Tempo ♩＝200+

Other ( )

**Appendix A:** Music Preference Checklist

## Music I'd like to listen to on a space travel

**1) Please choose your favorite genres form the list below. (You may select multiple.)**

Popular Songs / Hip Hop / Rock / Anime Songs / Vocaloid / Western Music / Classical / Jazz / Alternative / Blues / Country / Dance / R&B / Enka / Electronic / Soundtrack / Metal / K-POP / Latin

Other ( )

**2) Please select your preferred era (release period).**

～ 1970s / 1980s / 1990s / 2000s / 2010s / 2020s ～

**3) Please choose your preferred mood or tone of music. (You may select multiple.)**

Bright / Dark / Bittersweet / Fun / Humorous / Lively / Calm / Cool / Intense / Sad / Grand / Noisy / Warm / Sweet / Beautiful / Complex / Clear / Clean / Rebellious / Mature / Light & Rhythmical

Other ( )

**4) Please choose your preferred tempo (beats per minute).**

Slow Tempo ♩＝60–90

Medium Tempo (moderate speed) ♩＝100–110

Up-Tempo ♩＝110–140

Very Fast Tempo ♩＝150–200

Extremely Fast / Aggressive Tempo ♩＝200+

Other (